\documentclass[12pt]{article}

 \usepackage{paralist} 
     \usepackage{placeins} 
     \usepackage{amsmath,amsthm,amssymb}
     \usepackage{epsfig}
     \usepackage{psfrag}
     \usepackage{tabls}
     \usepackage{supertabular}
     \usepackage[square,comma,numbers]{natbib}
     \usepackage[top=2.5cm, bottom=2.5cm, left=2.5cm, right=2.5cm]{geometry}
\usepackage{indentfirst} 
\usepackage{colortbl} 
\definecolor{niceyellow}{rgb}{0.98,0.92,0.73}
\usepackage{fancyhdr}
\usepackage{fancybox}
\usepackage{hyperref} 
    \newcommand{\myfecha}{\today}

\begin{document}
\begin{center}
\Large{\bf  
The math must be right: Comment on 
``Dimensional analysis, falling bodies, and the fine art of 
not solving differential equations'' by C. F. Bohren [Am. J. Phys. {\bf 72},
534--537 (2004)]}
\end{center}
\begin{flushleft}
{\bf Sergio Rojas}\\
srojas@usb.ve \\
{Physics Department, Universidad Sim\'{o}n Bol\'{\i}var,
Ofic. 220, Apdo. 89000, Caracas 1080A, Venezuela.}
\end{flushleft}

{\bf Abstract}.
we find 
it necessary to 
advise 
the interested and active instructor of Physics on the wrongness of
some computations in the aforementioned article. Surprisingly, the 
Journal refuses to even publish an erratum on the paper, which naturally
brings to mind the question of the number of published papers 
requiring corrections.

\bigskip

\setlength{\parindent}{0.75cm}

In discussing the usefulness of dimensional analysis for performing
computations in physics, Bohren ended his interesting and inspiring article
pointing out that
``Unfortunately, because of the sheer drudgery of solving equations 
physical interpretation often is an afterthought instead of 
occupying pride of place, as it does in 
dimensional analysis.''\cite{Bohren:2004} In reply to this point of view
and favoring that in the teaching of physics courses, instructors 
do need to reinforce  the idea
that developing a taste for solving equations is not only  an 
integral part but also a 
very important
aspect in the understanding of physics, one could cite a recent 
letter
where the author regrets 
having lost the chance of wining the Nobel Prize because of
following 
John Wheeler's advice to ``never calculate without first knowing
the answer.''\cite{Goldman:2010}

In the spirit of not following Wheeler's advice, we redid some of Bohren's
computations in section II of his article. Starting from his equation
(10), one can perform a Taylor's expansion on the right hand side
of the expression
$\tau \sqrt{2g/R} = cosh^{-1}(1/(1-2(h/R))$ to obtain,
\begin{equation}
\tau = \sqrt{\frac{2h}{g}}\left(1 + \frac{5}{6}(\frac{h}{R}) 
                          + \frac{43}{40}(\frac{h}{R})^2
                          + \frac{177}{112}(\frac{h}{R})^3 + \cdots \right).
\label{eq:1}
\end{equation}
This result contains the right numerical factor for the leading error 
term $\epsilon = \frac{5}{6}(\frac{h}{R})$, as computed by equation (2)
in the article. Instructors might find it instructive to compare the 
result of equation 
(\ref{eq:1}) with
the one leading to Bohren's wrong result expressed in his equation (11).
It is obtained by
performing a Taylor's expansion on both side of his equation (10).

To find out about the correctness of the result given by our equation 
(\ref{eq:1}) one can resort to the exact solution of this problem.
Before doing that, let's mention
that the posed problem (i.e. 
the falling of a point particle in the gravitational field of a uniform
mass distribution spherically symmetric) is a completely 
solvable Newtonian mechanics problem,
which solution only requires the knowledge of some integrals and 
the chain rule 
for derivatives\cite{Rojas:2010ejp:b,Mungan:2009}. Nevertheless, this problem
is missing from the list of illustrative examples in practically
most commonly used textbooks. The exact solution 
can be written in the form,

\begin{equation}
\tau = \sqrt{\frac{L^3}{2G M}}\left[ \frac{\sqrt{\frac{L}{x}-1}}{\frac{L}{x}}
   + tan^{-1}\left(\sqrt{\frac{L}{x}-1}\right) \right],
\label{eq:2}
\end{equation}
where both $L$, the position at which 
the
point particle is released from rest at $\tau = 0$,
 and $x$, the position of the
point particle at any later time $\tau$, are measured from the center of 
the uniform mass distribution $M$ having spherical symmetry of radius $R$, 
and $G$ is the gravitational constant.

Now, by taking $L = R + h$ and $x=R$,
equation ($\ref{eq:2}$) can be written in the form,
\begin{equation}
\tau = \sqrt{\frac{R^3(1+h/R)^3}{2G M}}
\left[ \frac{\sqrt{\frac{h}{R}}}{1 + \frac{h}{R}}
   + tan^{-1}\left(\sqrt{\frac{h}{R}}\right) \right].
\label{eq:3}
\end{equation}
Taylor's expansion of this expression yields (here 
$g = G M/R^2$, is the local gravitational
constant)
\begin{equation}
\tau = \sqrt{\frac{2h}{g}}\left(1 + \frac{5}{6}(\frac{h}{R}) 
                                  - \frac{1}{40}(\frac{h}{R})^2
                          + \frac{9}{560}(\frac{h}{R})^3 + \cdots \right),
\label{eq:4}
\end{equation}
thus confirming the result reported via equation (\ref{eq:1}) to leading terms,
and providing corrections for additional higher order terms.  

In presenting the previous computations, the active instructor should not miss
the chance of talking about the fact
that numerical factors
are important when deciding on competing theories explaining physical
phenomena. 
A remarkable example is
provided by two  models competing to explain the drag $C_D$ on spherical
bodies falling in a resistive medium:\cite{VanDike:1975,VeyseyGoldenfeld:2007}
 the Oseen approximation,
equation (\ref{eq:5a}), 
which was discarded in favor of the correct model provided by
Proudman and Pearson, equation (\ref{eq:5b}),

\begin{subequations}
\label{eq:5}
\begin{gather}
C_D = \frac{6\pi}{R_e}\left(1 +  \frac{3}{8}R_e 
                              - \frac{19}{320}R_e^2
                          + \frac{71}{2560}R_e^3 + \cdots \right),
\label{eq:5a} \\
C_D = \frac{6\pi}{R_e}\left(1 +  \frac{3}{8}R_e 
                              + \frac{9}{40}R_e^2 ln(R_e) 
                              + \cdots \right),
\label{eq:5b} 
\end{gather}
\end{subequations}
where $R_e$ represents the Reynolds number (introduced in section VI
of Bohren's article). 
Certainly, experiments help in choosing the right model based on
 the numerical answer provided by each model as compared by the
respective measurement. Additional examples illustrating this
idea can be
drawn from
the tests that are applied to competing theories which try to explain
the perihelion shift of the planet Mercury and the deflection of light
by the sun. Einstein's General Relativity theory is the winner on the
basis that its predictions are in agreement with the results of 
increasingly accurate experiments\cite{Schutz:1985}.   

Let's finish this comment by pointing out that, aside from the
above discussion,  the examples
presented in Bohren's article \cite{Bohren:2004} 
could be applied to help students to practice what they have learned
in their math courses. Let's recall that it is in physics courses where
students can strengthen their quantitative reasoning 
skills\cite{Rojas:2010rmf,RojasS:2009,RojasS:2008}. 
Paraphrasing Heron and Meltzer,
learning to approach problems in a systematic way starts from 
learning the interrelationships among conceptual knowledge,
mathematical
skills and logical reasoning.\cite{HeronMeltzer:2005}
In physics, this necessarily requires the teaching of a good
deal of mathematical computations, and dimensional analysis could
be used to guide physical intuition.


\providecommand{\newblock}{}

\label{LastPage}

\end{document}